\RequirePackage{fix-cm}
\documentclass{svjour3}                     

 \usepackage[letterpaper, margin=1in]{geometry}

\usepackage{graphicx}
\usepackage{pgfplots}
\pgfplotsset{compat=newest}
\usepackage[utf8]{inputenc}

\usepackage[]{subcaption}
\usepackage{natbib}

\usepackage{amssymb}
\usepackage{verbatim}
\usepackage[titletoc,toc,title]{appendix}
\usepackage{xcolor}
\usepackage[export]{adjustbox}
\usepackage{multirow}
\usepackage{upgreek}
 \usepackage[switch]{lineno}
\newcommand{\mkf}{\textcolor{black}}
\newcommand{\myline}[1]{{\color{#1}${\rule[2pt]{.35cm}{2pt}}$}}

\begin{document}

\title{A single-camera, 3D scanning velocimetry system for quantifying active particle aggregations \thanks{Funding for this project was generously provided by the Gordon and Betty Moore Foundation}
}

\titlerunning{Scanning PIV System for quantifying migrating zooplankton aggregations}        

\author{Matt K. Fu         \and
        Isabel A. Houghton\and    
        John O. Dabiri 
}


\institute{M. K. Fu and J. O. Dabiri \at
              California Institute of Technology\\
              GALCIT \\ 
              1200 E California Blvd. \\
              MC 105-50 \\
              Pasadena, CA 91125\\
              \email{mkfu@alumni.princeton.edu}  \\
              \and
              I. A. Houghton \at
              Sofar Ocean\\
                Pier 50, Shed B, Bulkhead Office\\
                San Francisco, CA 94158\\
}

\date{Received: date / Accepted: date}

\maketitle

\begin{abstract}
A three-dimensional (3D) scanning velocimetry system is developed to quantify the 3D configurations of particles and their surrounding volumetric, three-component velocity fields. The approach uses a translating laser sheet to rapidly scan through a volume of interest and sequentially illuminate slices of the flow containing both tracers seeded in the fluid and particles comprising the aggregation of interest. These image slices are captured by a single high-speed camera, encoding information about the third spatial dimension within the image time-series. Where previous implementations of scanning systems have been developed for either volumetric flow quantification or 3D object reconstruction, we evaluate the feasibility of accomplishing these tasks concurrently with a single-camera, which can streamline data collection and analysis. The capability of the system was characterized using a study of induced vertical migrations of millimeter-scale brine shrimp (\textit{Artemia salina}). Identification and reconstruction of individual swimmer bodies and 3D trajectories within the migrating aggregation were achieved up to the maximum number density studied presently, $8 \, \times\,10^5$ animals per $\textrm{m}^3$. This number density is comparable to the densities of previous depth-averaged 2D measurements of similar migrations. Corresponding velocity measurements of the flow indicate that the technique is capable of resolving the 3D velocity field in and around the swimming aggregation. At these animal number densities, instances of coherent flow induced by the migrations were observed. The accuracy of these flow measurements was confirmed in separate studies of a free jet at $Re_D = 50$.
\keywords{First keyword \and Second keyword \and More}
\end{abstract}

\section{Introduction}
\label{sec:intro}
Turbulent flows containing dispersed particles are a common feature in many environmental and industrial processes. The particles within these flows include both passive phases such as solid particles \citep{Balachandar2010}, bubbles \citep{Rensen2005,Risso2018}, and droplets \citep{Aliseda2021}, as well as `active' phases such as swimming zooplankton \citep{Jumars2009}. At sufficiently large particle volume fractions ($10^{-6} \leq \Phi_v \leq 10^{-3}$), the presence of the particles creates unique flow dynamics associated with the two-way fluid-particle coupling that are distinct from single-phase turbulence \citep{Elghobashi1994}. Characterizing this two-way coupling requires accurately reconstructing the three-dimensional (3D) aggregations of particles and the turbulent flow field in which they are dispersed. Because the fluid-particle interactions are 3D and occur over a wide range of spatiotemporal scales, there are many challenges to measuring them experimentally. These challenges are exacerbated in denser aggregations where there are larger numbers of particles and interactions that need to be tracked and quantified \citep{Bourgoin2014}.

Biologically generated turbulence is an emerging topic whose study is currently limited by an inability to quantify the flow within aggregations of swimming plankton. The turbulence created by these aggregations remains a poorly understood, and potentially underrepresented, source of scalar transport and ocean mixing \citep{Kunze2019}. Though the eddies created by an isolated swimmer are comparable to that of the individual organism, the larger length scales associated with the aggregations of swimmers have the potential to introduce mixing scales relevant to the surrounding water column. Recent laboratory studies of millimeter-scale brine shrimp (\textit{Artemia salina}) aggregations using two-dimensional (2D) flow measurement techniques have shown that induced migrations could generate aggregation-scale mixing eddies through a Kelvin-Helmholtz instability \citep{Wilhelmus2014} with effective turbulent diffusivities several orders of magnitude larger than molecular diffusion alone \citep{Houghton2018,Houghton2019}. Though the potential for enhanced mixing is substantial, direct measurements of enhanced turbulent dissipation and mixing in lakes and the ocean due to vertical migrations have been less conclusive \citep{Noss2014,Simoncelli2018,Kunze2019}. Parameterizing the precise conditions and mechanisms that lead to enhanced mixing remains an active area of research \citep{Wang2012,Wang2015,Ouillon2020,More2021}.

There are numerous efforts to develop volumetric velocimetry techniques capable of resolving the unsteady flow field in addition to the morphology and kinematics of a single swimming organism. A common technique for volumetric, three-component (3D-3C) velocity measurements is tomographic particle image velocimetry (Tomo-PIV), which has been used extensively for investigations of aquatic locomotion, including the propulsive mechanisms of fish \citep{Gemmell2019} and pteropods \citep{Adhikari2016}. A key requirement for Tomo-PIV is employing four or more cameras to provide sufficient viewing angles for the tomographic reconstruction of both tracer particles used for flow quantification and swimmer bodies. Though there have been significant advancements in the resolution of Tomo-PIV for velocity quantification, most notably, through the `Shake-the-Box algorithm' of \citet{Schanz2016}, accurately reconstructing active or passive particles with complex, three-dimensional shapes remains challenging. 
One common approach to body reconstruction is to compute a visual hull based on the projection of an object onto multiple camera viewpoints \citep{Adhikari2012}. This method can overestimate the body size and obscure complex or rounded body geometries. While these shortcomings can be moderated by prescribing additional constraints to the body morphology or kinematics, such an approach typically requires \textit{a priori} knowledge of the behavior of the dispersed phase \citep{Ullah2019}. Despite these advancements, accurately reconstructing dense aggregations of particles, especially those with complex morphology, remains elusive. 


Beyond Tomo-PIV, several alternative 3D-3C techniques have been proposed for marine swimming quantification, including plenoptic imaging \citep{Tan2020}, synthetic aperture particle image velocimetry \citep{Mendelson2015,Mendelson2018}, defocusing digital particle image velocimetry (DDPIV) \citep{Pereira2002, Troutman2018}, and 3D digital holography \citep{Gemmell2013}. Though all of these techniques have been demonstrated on individual swimmers, few are suitable for object reconstruction, and none have been successfully deployed to reconstruct dense configurations of swimmers and tracer particles in 3D. 
%

Here, we present a 3D scanning system to reconstruct configurations of vertically migrating swimmers and quantify their surrounding 3D-3C velocity field. Several scanning systems have been developed in recent years for a variety of applications, including 3D-3C velocity measurements \citep{Hoyer2005,Brucker2013,Lawson2014,Ni2015,Kozul2019} and 3D object reconstruction of translucent organisms \citep{Katija2017a, Katija2020a} and structures \citep{Buehler2018}. The 3D scanning system in the present study is conceptually similar to those existing systems but is used to simultaneously quantify the locations and organizations of the swimmers and their surrounding flow field. The approach relies on a laser sheet that rapidly and repeatedly scans through a volume of interest, sequentially illuminating image slices of flow tracer particles and organism cross-sections. The images are captured by a single high-speed camera, encoding detailed information about the third spatial dimension within the image time-series. Repeated scanning creates a series of image volumes consisting of swimmer bodies and tracer particles. Due to their large size relative to the tracer particles, the swimmer bodies can be identified and tracked over time. Similarly, the velocity field in the vicinity of the swimmers is determined via localized 3D cross-correlations of consecutive tracer particle images.

The capabilities of the technique are demonstrated by scanning induced vertical migrations of brine shrimp (\textit{Artemia salina}). We demonstrate that the 3D position, orientation, and morphology of individual \textit{A. salina} can be faithfully reconstructed, even at large animal number densities up to $8 \times 10^5$ animals per $\textrm{m}^{3}$, the high end of previously reported brine shrimp number densities in the literature \citep{Houghton2019}. We then show selected examples in which a coherent, large-scale induced flow is resolved by the measurement technique. While the appearance of large-scale induced flow was not observed during each migration, the present results demonstrate the ability of the measurement technique to capture those dynamics when they do occur. Lastly, the outlook for the technique is discussed with suggested technical improvements to the system design. 

\section{Scanning 3D Image Reconstruction System}
\label{sec:scanning}


\subsection{Imaging Hardware and Procedure}
\label{sec:scanning:hardware}

\begin{figure*}[ht!]
    \centering
        \includegraphics[width = \textwidth]{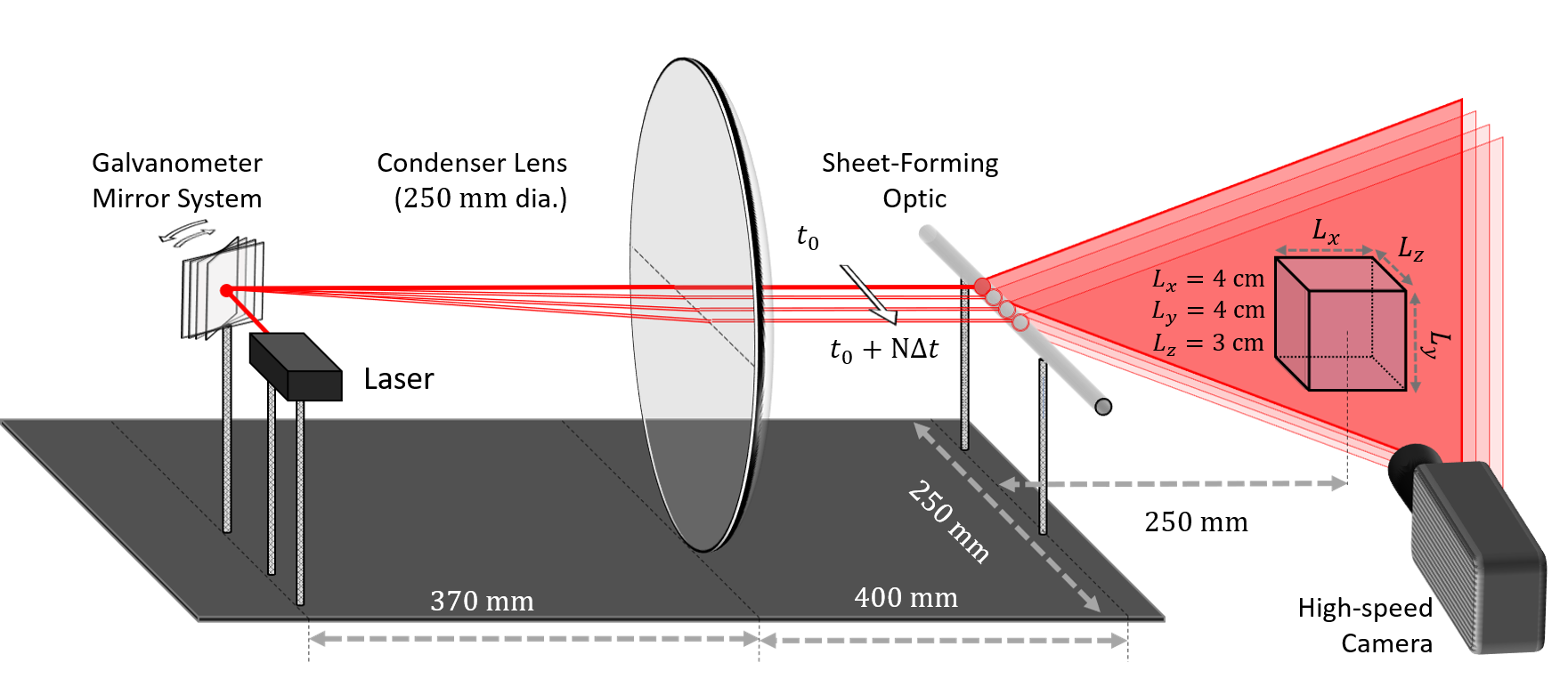}
        \caption{Diagram of the scanning system. A mirror-mounted galvanometer (left) deflected the imaging laser along the scanning direction. A condenser lens (center-left) collected the angled beams to create displaced but parallel beams. The beams were then formed into scanning sheets by a long sheet forming optic (center-right). The imaging volume (right) was repeatedly scanned to sequentially illuminate slices of tracers and particles. The image slices were captured by a high-speed camera (lower-right) and then stacked to form 3D image volumes.}
        \label{fig:isoView}
\end{figure*}

The design of the scanning system, shown in Figure \ref{fig:isoView}, was similar to the system of \citet{Lawson2014}. Illumination for the scanning was provided by a 671 nm continuous wave laser (5-Watt Laserglow LRS-0671 DPSS Laser System). This wavelength of light ensured that the brine shrimp exhibited no phototactic response to the imaging light. Additionally, the laser beam had only a single Transverse Electric Mode (i.e., near TEM$_{00}$ or quasi-Gaussian beam) to minimize imaging artifacts along the scanning dimension due to the beam shape. 

The laser beam was angled along the scanning dimension of the imaging volume by a mirror with a broadband dielectric coating (-E02) mounted on a single-axis galvanometer (Thorlabs GVS211/M). The angular range (max $\pm 20^\circ$) and bandwidth ($65$ Hz square wave at $50\%$ full travel) of the galvanometer were comparable to other scanning systems in the literature that rely on scanning optics such as rotating polygonal mirrors \citep{Hoyer2005,Brucker2013} or piezo-electric mirrors \citep{Ni2015}. An analog voltage signal from an arbitrary function generator (Tektronix AFG3011C) controlled the tilt of the mirror, which determined the position and scanning rate of the laser.
The angled beams were collected by a 250 mm dia. condenser lens (370 mm back focal length), realigning them into parallel trajectories displaced along the scanning direction. These scanning beams were then converted into scanning sheets by a sheet forming optic that spans the depth of the imaging volume, such as a glass cylinder. The size of the condenser lens and the length of the sheet forming optic determined the maximum distance over which the beams could be collected and aligned. 
By employing a condenser lens with a relatively large focal length, the amount of mirror rotation necessary to deflect the beams over the entire depth of field was contained to just a few degrees ($\pm 1.2^\circ$ in the present study). Here, the galvanometer was driven with a sawtooth wave to repeatedly scan the imaging volume with a constant forward scanning speed that filled approximately $94\%$ of the scanning period. The remaining $6\%$ of the scanning period was spent on the backward scan to reset the mirror position for the next imaging period. The accuracy of the scanning rate was limited by the repeatability of the galvanometer ($0.07\%$ for $30\, \mu$rad beam angle repeatability).


By rapidly scanning a laser sheet along the sheet-normal axis, 1 millimeter-thick image slices throughout the depth of the interrogation volume were sequentially illuminated and captured by a high-speed camera. By ensuring that the scanning period was considerably faster than the flow time scales (e.g. laser translation speed 30 times faster than the animal swimming speed in the present experiments), the recorded images could encode spatial information about the scanning dimension within the image time-series. The image sequences were stacked to construct volumetric (3D) images of the quasi-static tracers and larger active or passive particles, such as the swimmers of present interest. Periodic scanning of the interrogation volume facilitated tracking of the particles and tracers over time. 

\subsection{Imaging Acquisition \& Calibration}
\label{sec:scanning:calibration}

The scanned images were captured with a high-speed camera (Photron FASTCAM SA-Z) equipped with a fixed focal length macro lens (Micro-NIKKOR 105 mm with a 36 mm extension tube) at $1024\,\times\,1024\,\textrm{px}^2$ resolution. The image acquisition rate was matched to the scanning speed such that the displacement of the laser sheet between each frame was approximately the same size as the mean pixel resolution (i.e., $40 \,\mu$m). This fine depth-wise sampling allowed the raw image volume to have a nearly isotropic voxel size. Both the f-number ($f/22$) and working distance (approximately $0.4-0.5$m) were iteratively tuned to ensure that the entirety of the imaging volume was within the depth of field ($3$ cm) and each scanned image was in sharp focus. 

\begin{figure}[ht]
    \centering
    \begin{subfigure}[b]{.4\textwidth}
    \includegraphics[width = \textwidth]{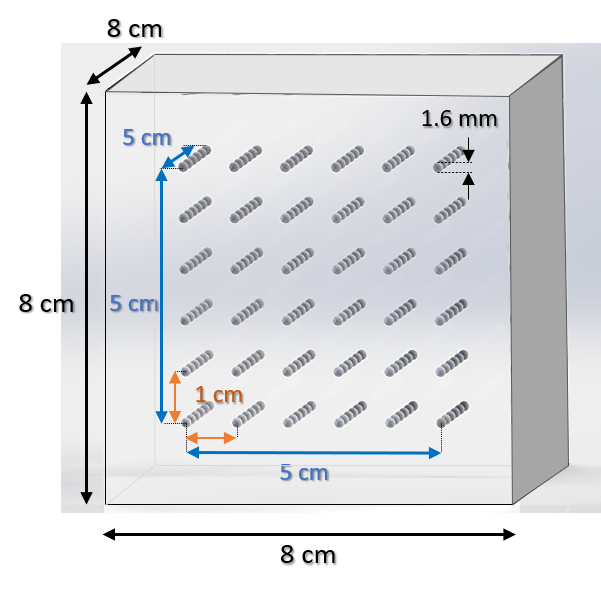}
    \caption{} 
    \label{fig:cube_pic}
    \end{subfigure}
        \begin{subfigure}[b]{.48\textwidth}
    \includegraphics[width = \textwidth,trim= 0cm 0cm 1cm 1cm, clip]{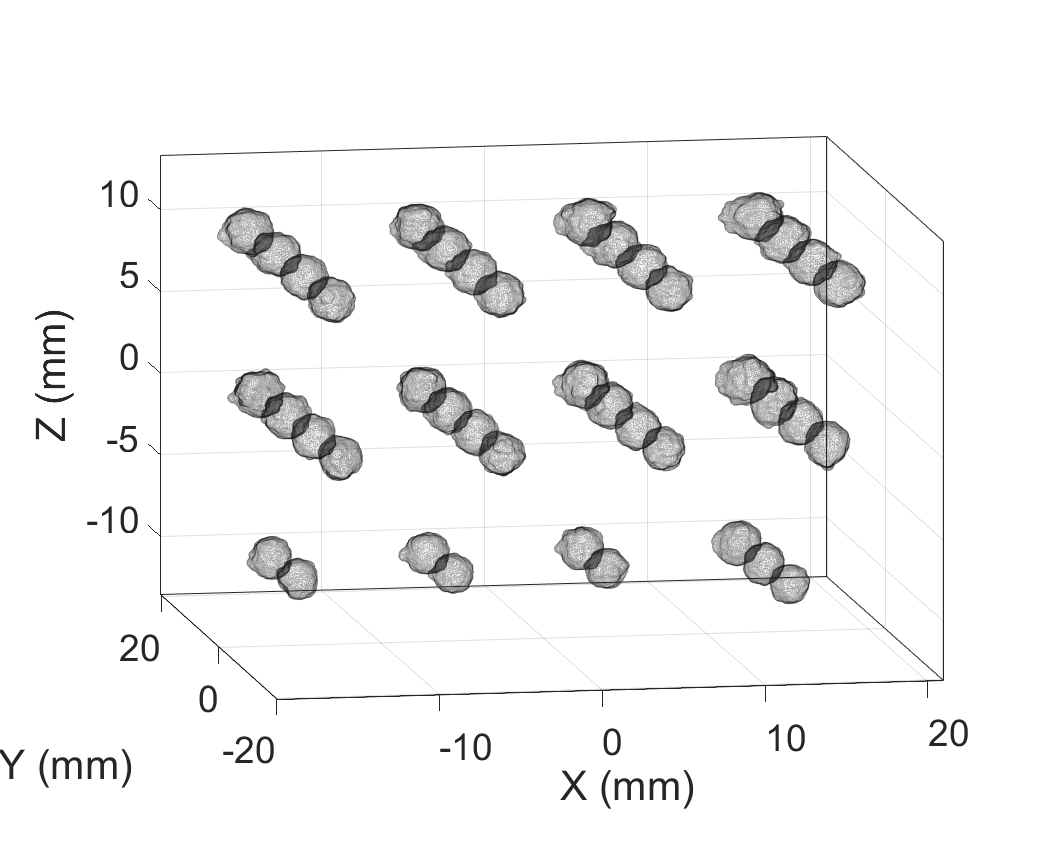}
    \caption{} 
    \label{fig:cube_scan}
    \end{subfigure}
    \caption{Images of the calibration cube target. (a) CAD Rendering of the 3D calibration cube. The $6\times6\times6$ array of spheres can be seen in the center of the $8\times8\times8$ $\textrm{cm}^3$ cube. (b) Rendered scan of the calibration cube from the 3D scanning system. Some spheres are not reconstructed in the image due to low scattering intensity. }
        \label{fig:cube}

\end{figure}

A custom 3D calibration target (UCrystal) was fabricated to calibrate the imaging volume and account for the \mkf{$7\%$} change in magnification along the scanning depth. The target, shown in Figure \ref{fig:cube_pic}, comprised an $8  \;\rm{cm}\;  \times\;  8\;\rm{cm}\; \times\; 8\;  \rm{cm}$ crystal cube internally laser engraved with a 3-dimensional grid of $1.6 \; \rm{mm}$ diameter spherical shells. The shells were evenly spaced $1 \; \rm{cm}$ apart in each direction to form a $6 \; \times\; 6 \;\times\; 6$ cubic array ($5 \; \rm{cm} \; \times \; 5 \;\rm{cm} \; \times 5\; \rm{cm}$), which was centered within the crystal. The spot size of the laser engraver used to raster the spherical shells was approximately 100 $\mu \rm{m}$. The cube was suspended at the center of the imaging volume and \mkf{aligned with the imaging coordinate system} to ensure that the laser sheet was not deflected by refraction inside the cube. 

Calibrating the imaging system involved scanning the calibration cube with the laser sheet and capturing the 2-D image slices with the high-speed camera. The images collected over each period were stacked to form a single 3-D image volume. Because the scanning was designed to create nearly isotropic volumes, minimal processing of out-of-plane dimension was necessary to render scanned objects. 

The raw image volumes were processed and analyzed using MATLAB's Image Processing Toolbox to reconstruct and locate the spherical targets. The image volume was median filtered ($7^3$ vx. stencil) and binarized with a global threshold based on the image histogram. Morphological area opening was then used to remove objects other than the calibration spheres, e.g., tracer particles and camera noise, from the binary image, leaving just the calibration spheres. Any holes within the binary images of the spheres were then filled. The centroids of the remaining spheres (shown in Figure \ref{fig:cube_scan}) were then used to calibrate the image volume. While all of the target spheres were scanned, not all of them were successfully reconstructed. This failure was most common in target spheres further from the camera as their scattered light could be obstructed by spheres in the foreground. 

By relating the centroids of the rendered spheres to the known dimensions of the calibration target, the voxels within the image volume could be mapped to 3D coordinates in physical space. The mapping between the two coordinate systems was calculated using the MATLAB \verb+estimateCameraParameters+ function.

\subsection{Particle Segmentation}
\label{sec:scanning:processing:segmentation}


Just as the spheres were extracted from the calibration target, we do the same for the active/passive particles in an aggregation. Because the particles in this study, i.e., the swimmers, were significantly larger than the tracers, they could be identified and segmented within the image volume by size. This segmentation process was accomplished by filtering the raw images with a cubic Gaussian kernel (3 vx. stencil). The filtered images were then binarized with the method of Otsu \citep{Otsu1979}, which computes a global threshold based on the image histogram. Tracers were removed from the binary image by filtering out objects smaller than 8000 connected voxels through morphological area opening. This 8000-voxel threshold was found to work satisfactorily for the specific imaging parameters in this study. Depending on the application and object size distribution, alternative segmentation techniques, such as the 3D analogs of those reviewed by \citet{Khalitov2002}, may prove more robust. Connected components within the binary image were labeled as individual swimmer bodies. The centroids of each of the swimmer bodies were tracked over time to determine the swimmer trajectories. A mask for the tracer field was computed by morphologically dilating the binary image of the particles with a spherical structuring element (4 vx. radius).


\subsection{Velocity Field Registration}
\label{sec:scanning:processing:velocity}

With the particles comprising the aggregation segmented, the remainder of the image corresponding to the tracer field was then used to compute the volumetric, three-component velocity field by registering the local displacements of tracer particles between successive images. Each pair of image volumes was first masked using the binary images of swimmer bodies from the previous segmentation step. Each mask was applied to both images in the pair to ensure each frame had an identical mask and avoid correlations due to mask shifting. 

To resolve the local tracer displacement between consecutive images, we employed a modified version of the Fast Iterative Digital Volume Correlation (FIDVC) Algorithm of \citet{BarKochba2015}. This method could resolve large volumetric deformations between two images by conducting 3D cross - correlations on progressively refined interrogation windows to compute the local image displacement. 
First, the original images were divided into $64\,\times\,64\,\times\,64\,\textrm{vx}^3$ windows with $50\%$ overlap. Each windowed image was weighted with the modular transfer function of \citet{Nogueira2005} to stabilize the spatial frequency content. The 3D, voxel displacement between the two images was determined to the nearest integer voxel by finding the local maximum of the cross-correlation function between the two windows.  Sub-voxel resolution for the displacement was then achieved by first conducting a least-squares fit with a 3D Gaussian function to the $5^3$ voxel neighborhood around the peak value in the cross-correlation function. The sub-voxel displacement was then determined by solving for the local maximum of the resulting fit. 

Displacement vectors with correlation coefficients below a certain threshold ($\leq0.01\%$ of the maximum correlation) or within the image mask were rejected and replaced with interpolated values. The displacement field was then filtered with the tunable low pass convolution filter of \citet{Schrijer2008} to improve the iterative image deformation, and all nonphysical outliers were removed via a universal median test \citep{Westerweel2005}. Both image volumes were then symmetrically deformed by a tri-cubic interpolation scheme using the MATLAB \verb+griddedInterpolant+ function. \mkf{The root mean square (RMS) deviations between the two images before and after deformation were computed and their ratio was used as a convergence metric. When the RMS deviation ratio was reduced to less than $0.1$, the window size was refined for the next iteration. The iterative deformation process was repeated until the minimum window size ($32\,\times\,32\,\times\,32\,\textrm{vx}^3$ with $75\%$ overlap) was reached and the final RMS ratio was less than $0.2$. These convergence criteria were found to provide an acceptable balance between accuracy and computation times for the images analyzed in this study and typically required 7 iterations to achieve convergence.} All Fast Fourier Transforms (FFTs) and sub-voxel estimation operations were executed with the MATLAB Parallel Computing Toolbox on two NVIDIA Quadro RTX5000 GPUs with double precision. This GPU variant was benchmarked against the original FIDVC code \citep{BarKochba2015} with agreement found in all cases up to single precision.

\section{Induced Vertical Migrations of \textit{Artermia Salina}}
\label{sec:exper}
\begin{figure*}[ht]
    \centering
    \begin{subfigure}[b]{0.3\textwidth}
        {\includegraphics[trim = {0 0 0cm 0}, clip, height = 8cm,left]{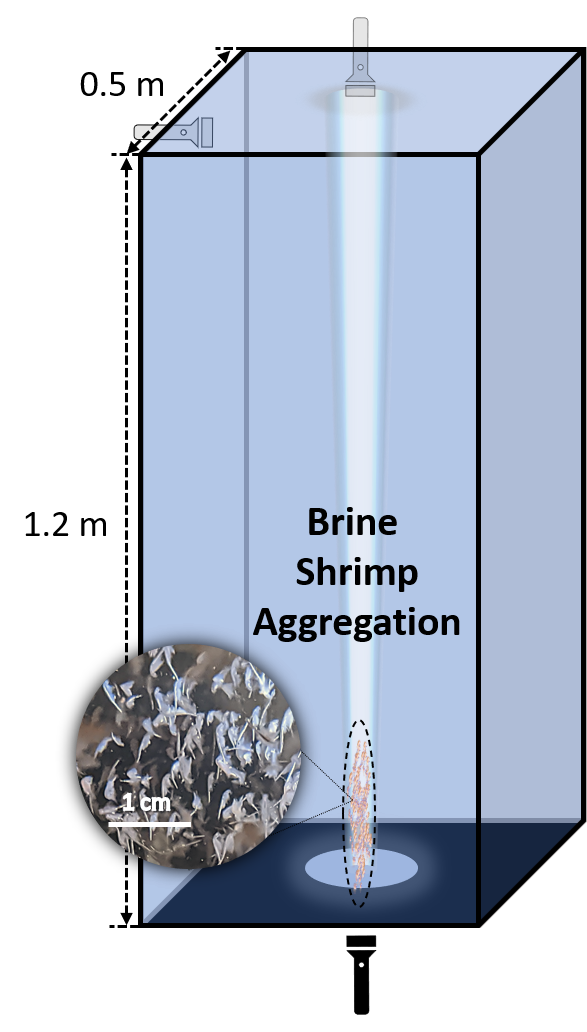}}
        \caption{}
        \label{fig:tank1}
    \end{subfigure}
    \begin{subfigure}[b]{0.3\textwidth}
        {\includegraphics[trim = {0 0 0cm 0}, clip, height =8cm,right]{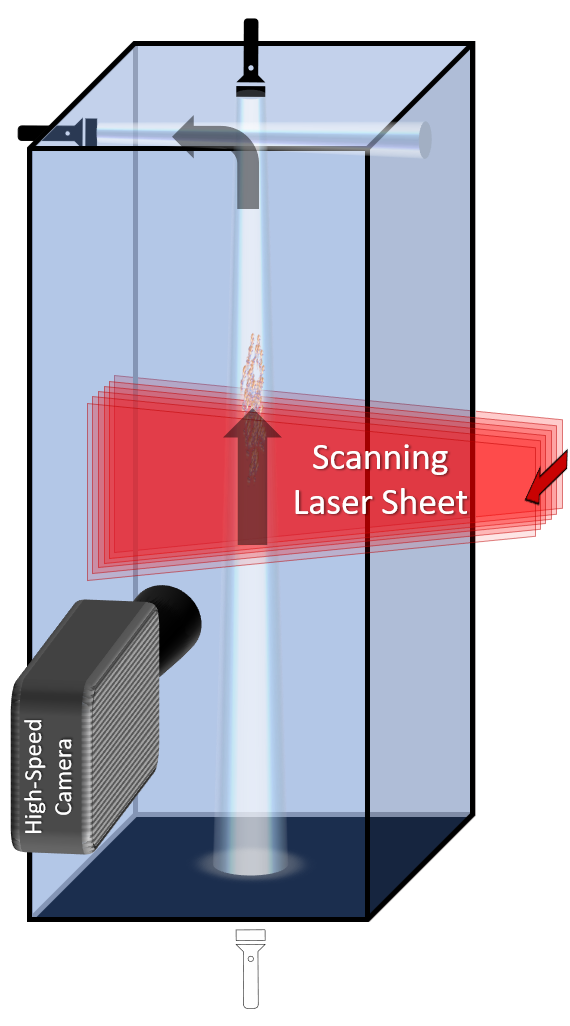}}
        \caption{}
        \label{fig:tank2}
    \end{subfigure}
    \begin{subfigure}[b]{0.3\textwidth}
        {\includegraphics[trim = {0cm 0 0cm 0}, clip, height =8cm,center]{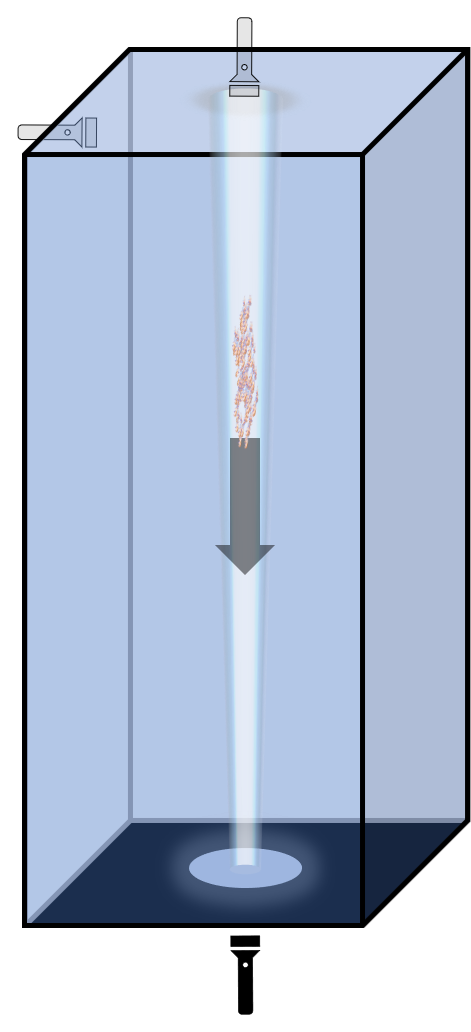}}
        \caption{}
        \label{fig:tank3}
    \end{subfigure}
    \caption{Illustration of the \textit{Artemia salina} vertical migration and image acquisition. (a) The brine shrimp were collected at the bottom of the 1.2 m tall tank with an upward projecting LED light stimulus (PeakPlus LFX1000, 600 lumens). After the \textit{Artemia salina} reached the bottom, the tank was allowed to settle for at least 20 min. to ensure the fluid was quiescent. (b) An upward migration was induced with a similar downward projecting light stimulus at the top of the tank. An additional light was used to steer the brine shrimp horizontally along the free surface to the side of the tank, reducing the number of swimmers that would accumulate and obstruct the downward projecting light. The imaging volume was positioned along the  central axis of the tank approximately 40 cm below the tank free surface. (c) A downward migration was induced with the first upward-facing light to return the shrimp to the bottom of the tank.}
            \label{fig:tank}
\end{figure*}

To test the capability of the technique in capturing aggregation kinematics and associated fluid mechanics, we evaluated vertical migrations of brine shrimp (\textit{Artemia salina}) within a laboratory tank following the methodology of \citet{Houghton2018} and \citet{Houghton2019} and imaged the resulting flow. This application was selected due to the challenge that the animal number density presented to existing techniques. Additionally, the slow evolution of the migration was compatible with the achievable scanning rate of the current system ($\mathcal{O}$(1) sec). By leveraging the positive phototaxis of \textit{A. salina} towards sources of blue and green wavelengths of light, coordinated swimming of a brine shrimp aggregation could be directed up and down the height of a 1.2-meter tall vertical tank (see Figure \ref{fig:tank}). A collection of approximately $40,\!000\pm5,000$ animals (Northeast Brine Shrimp) was introduced to the tank for testing, corresponding to a tank-averaged abundance of $130,000\pm16,000$ animals per m$^{3}$. The brine shrimp had a typical body length of $5$ mm and a nominal swimming speed of $5$ mm/s. The tank was seeded with 13 $\mu \textrm{m}$ CONDUCT-O-FIL silver coated glass spheres (Potters Industries, Inc.) to facilitate imaging of the flow field. 

Before the migration, the animals were collected at the bottom of the tank using an upward facing light stimulus (PeakPlus LFX1000, 600 lumens) introduced through the transparent floor of the water tank. After the animals reach the bottom of the tank, the water was allowed to equilibrate for at least 20 minutes to ensure the fluid was quiescent. Due to the slight negative buoyancy of \textit{A. salina}, the animals were minimally active at the bottom of the tank. To trigger the upward migration, the light stimulus at the bottom of the tank was deactivated, and corresponding light stimuli at the top of the tank were activated. The first of these lights (PeakPlus LFX1000, 600 lumens) was directed down along the tank's central axis in a $5\pm2$ cm diameter column and served as the primary stimulus to draw the animals up towards the free surface. A second horizontal light (PeakPlus LFX1000, 600 lumens), located just below the free-surface, steered the animals along the free surface and away from the primary stimulus to prevent them from accumulating and obstructing the migration. The duration of the vertical migration, typically 5-6 minutes, extended from the triggering of the lights until the accumulated \textit{A. salina} began to obstruct the primary stimulus.

The 3D scanning system imaged the swimmer aggregation and tracers within in a $41\times41\times30$ $\textrm{mm}^3$ volume approximately 40 cm below the free surface. Throughout the vertical migration, scanning sequences were triggered at approximately 1-minute intervals to record a sequence of approximately $22,\!000$ images, corresponding to a minimum of 26 image volumes over a 5 second period. The duration of the scanning sequence was limited by the size of the camera internal buffer (32 GB), and the 1-minute interval between scanning events was dictated by the time necessary to fully transfer the images to an external hard drive. Following the migration, the animals were returned to the bottom of the tank using the light stimulus under the transparent floor of the water tank. The complete imaging volume specifications and scanning parameters can be found in Table \ref{tab:imaging_params_brineshrimp}.

\begin{table}[ht!]
\centering
\scriptsize
\begin{tabular}{lr|c|c}
\hline
&Parameter    &     Symbol         &        Value           \\
\hline \hline
\parbox[c]{1mm}{\multirow{11}{*}{\rotatebox[origin=t]{90}{Scanning Parameters}}}
&Field of View       &   $L_x \!\times\! L_y$  &   $41\,\times41\,\textrm{mm}^2\,$ \\
&Depth of Field      &   $L_z$    &   $30 \,\textrm{mm}$ \\
&Approx. Voxel Size  &   \rule{0.5cm}{0.5pt}     &  
$40\times40\times40\,\mu\textrm{m}^3$ \\
&Image Acq. Rate    &   $f_c$     &   $4,000$ fps\\
&Scanning Freq.    &   $f_s$     &   $5$ Hz\\
&Scanning Speed    &   $u_s$     &   $15\;\textrm{cm} \cdot \textrm{s}^{-1}$ \\
&Sheet thickness    &   $h$     &   $1\;\textrm{mm}$ \\
&Sheet separation    &   $\Updelta z$    &   $40\,\mu\textrm{m}$ \\
&Est. Seeding Density    &   $N_{V}$    &   $1.7\times 10^{-4}$ ppv\\
&No. Vols. per scan    &  \rule{0.5cm}{0.5pt}     &   26\\
&Sheet step size   &   $\Updelta z$    &   $40\,\mu\textrm{m}$ \\
\hline
\parbox[c]{1mm}{\multirow{3}{*}{\rotatebox[origin=c]{90}{Shrimp }}}
& Body Length    &   $\ell_c$    &   $5-10\,\textrm{mm}$ \\
& Rel. Swimming Speed    &   $U_{swim}$    &   $3\pm1 \;\textrm{mm/s}$ \\
& Tank-Avg. No. Density    &   $n$    &   $1.3 \pm 0.16\times 10^5$ $\textrm{m}^{-3}$ \\

\hline
&No. laser sheets  &   $N$     &   $750$ \\
&Non-dim Scan Speed    &   $u_s/U_{swim}$    &   30 \\
&Non-dim Sheet Width    &   $h/\Updelta z$    &   $20$ \\
\hline
\end{tabular}
\caption{Imaging parameters for brine shrimp migrations}
\label{tab:imaging_params_brineshrimp}
\end{table}






\section{Results}
\label{sec:results}

\subsection{Body Reconstruction and Tracking}
\begin{figure}[ht!]
    \centering
        \frame{\includegraphics[rotate =90,trim = {13.5cm 11cm 1.25cm 1cm}, width = 0.49\textwidth, clip ]{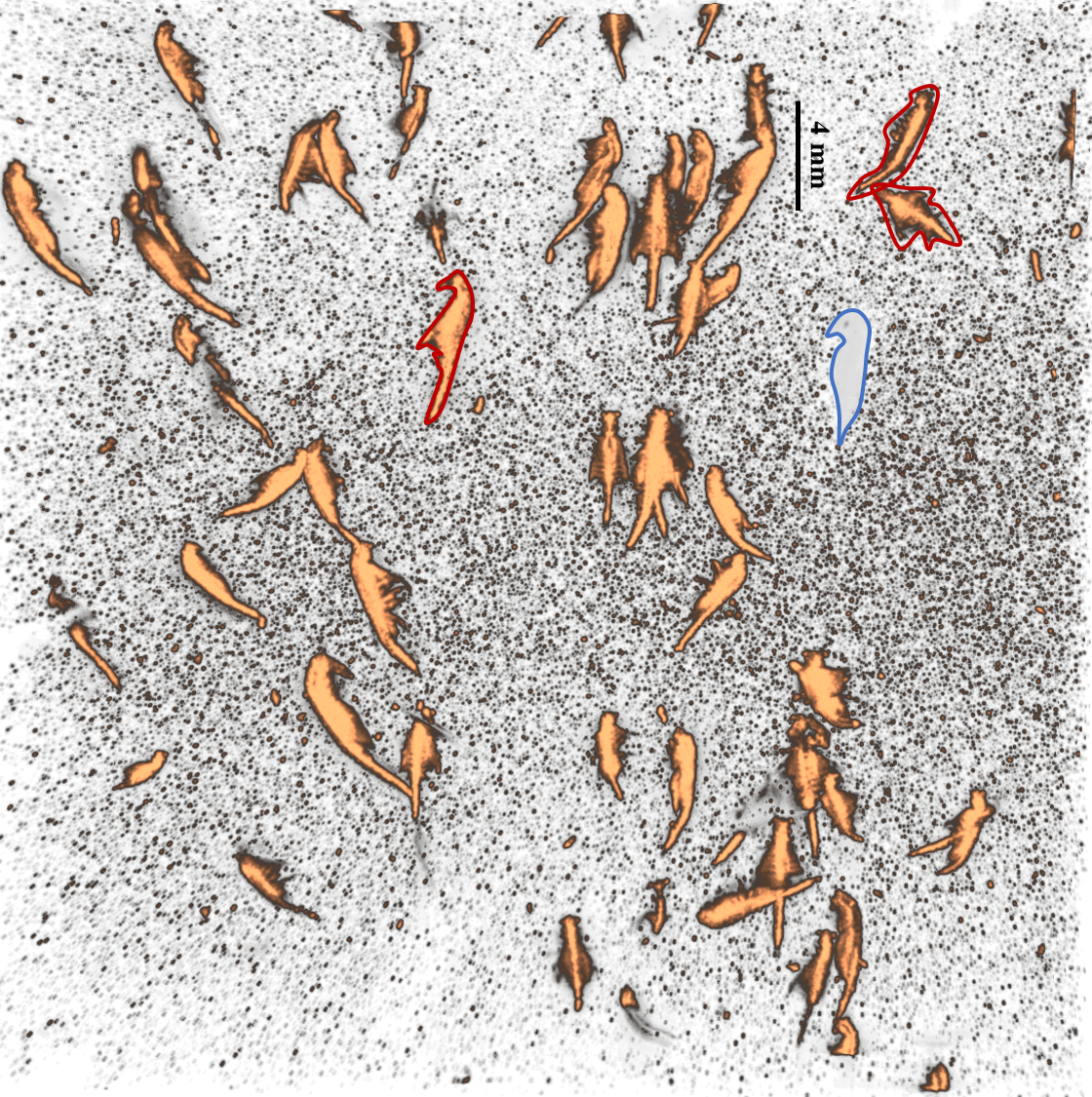}}
        \caption{Typical subsection of raw image volume of vertical migration from the camera perspective. Image intensity is inverted and colored for clarity. The shadowing effect from a foreground shrimp is outlined in blue. Reconstructed shrimp are shown in copper and outlined in red. Images of tracer particles can be seen as the dots interspersed throughout the image.}
        \label{fig:rawImVol}
\end{figure}
Following the procedure outlined in section \ref{sec:scanning:processing:segmentation}, individual shrimp bodies were segmented in the image volume to directly assess their number, location, and orientation. A representative portion of a raw image volume is shown in Figure \ref{fig:rawImVol} from the camera viewpoint. Due to the translucent nature of the shrimp bodies, light was readily scattered off the organisms, allowing them to be identified as large coherent objects within the 3D image amongst a field of smaller tracer particles. An example of two imaged shrimp are visible in the left side of Figure \ref{fig:rawImVol} with a copper coloring and outlined in red for clarity. While most of the details of the shrimp morphology are evident in the image, fine features such as the shrimp legs and tail are attenuated and blurred. Due to the nature of single-camera imaging, details of the shrimp bodies and particles can be obscured or altogether blocked by objects in the foreground. An example of this shadowing effect is shown outlined in blue on the right side of Figure \ref{fig:rawImVol}. Both the lack of visible particles and resemblance of the shadowed area to a shrimp silhouette indicated the presence of a shrimp located between the imaging volume and the camera.

\begin{figure*}[ht!]
\centering
\begin{subfigure}[b]{0.53\textwidth}
{\includegraphics[trim={0.4cm 0pt 1cm 0pt},width = \textwidth, clip]{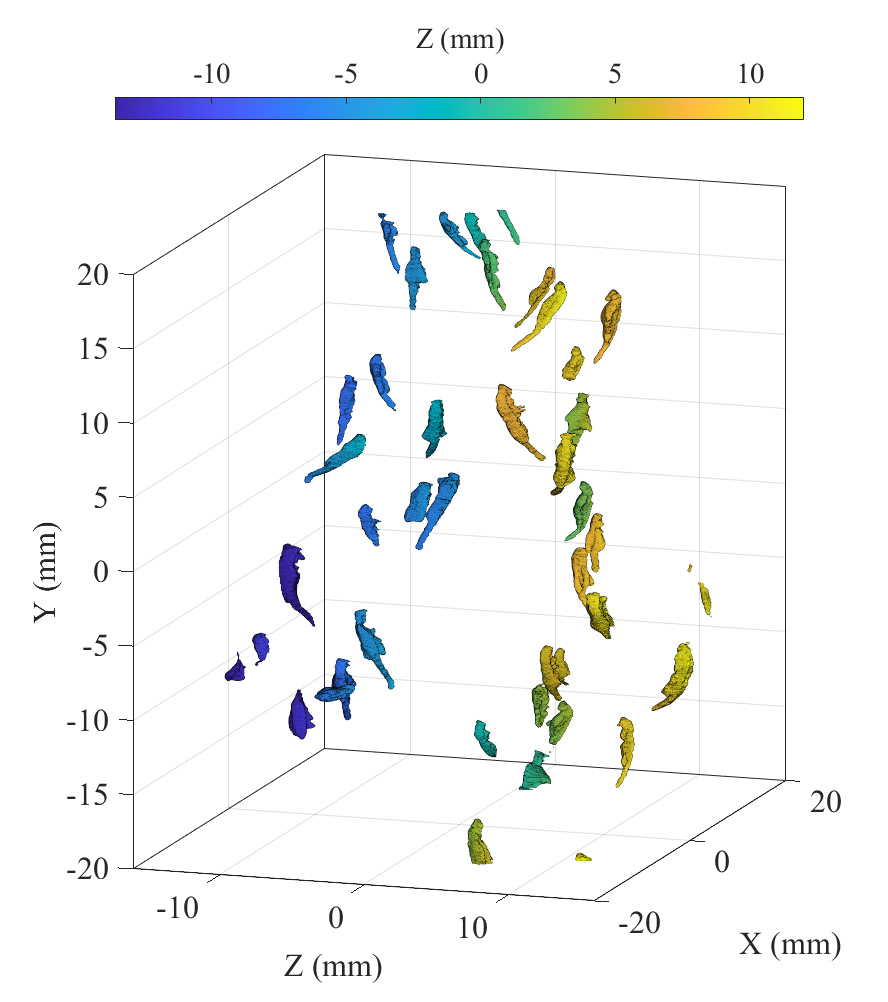}}
\caption{}
        \label{fig:3dmig}
\end{subfigure}
\centering
\begin{subfigure}[b]{0.46\textwidth}
\centering

\begin{picture}(100,100)
\put(0,0){{\includegraphics[trim={2cm 0pt 4cm 0pt},clip,height = 5.5cm]{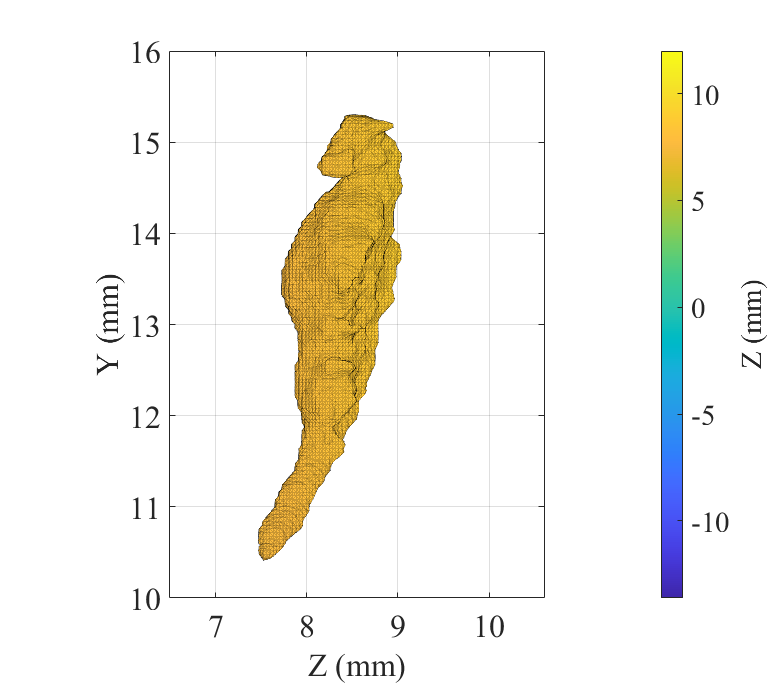}}}
\put(94,135){(b)}
\end{picture}\hspace{3mm}
\begin{picture}(100,100)
\put(0,0){{\includegraphics[trim={3.5cm 0pt 7cm 0pt},clip,height = 5.5cm]{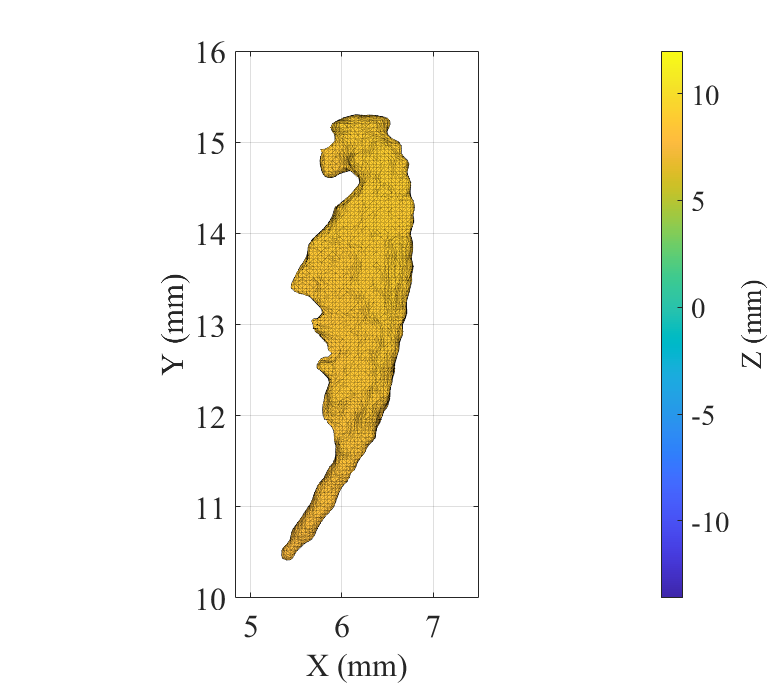}}}
\put(66,135){(c)}
\end{picture}
\vspace{0.8in}

\hspace{-3mm}
\begin{picture}(100,100)
\put(0,0){{\includegraphics[trim={2cm 0pt 5.5cm 0.5cm},clip,height = 5cm]{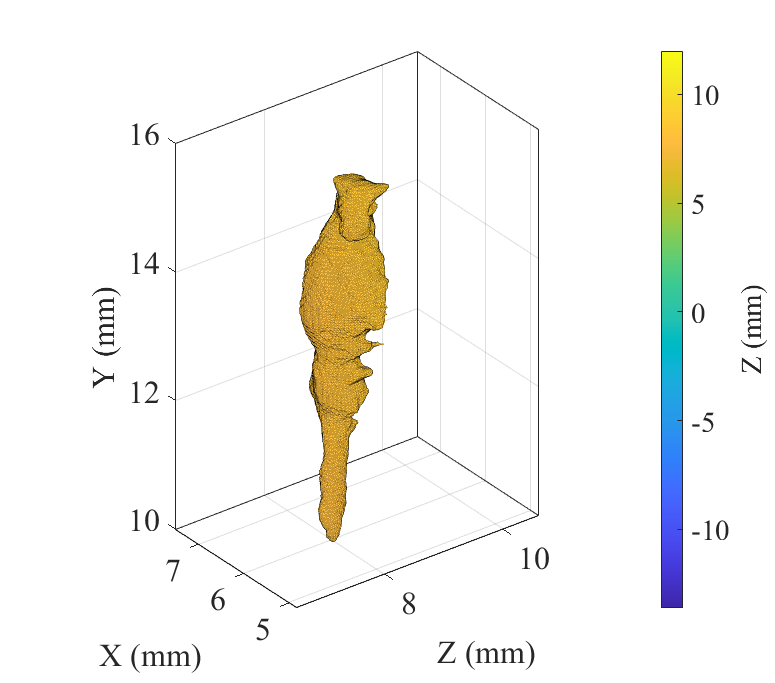}}}
\put(82,112){(d)}
\end{picture}
\begin{picture}(100,100)
\put(0,0){{\includegraphics[trim={-1cm -1cm 0 -1cm},clip,height = 5cm]{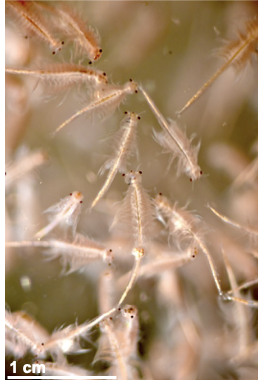}}}
\put(65,110){\textcolor{white}{(e)}}
\end{picture}


\end{subfigure}


\caption{Reconstructed swimmer bodies from an induced vertical migration. ($8\times 10^5$ animals per $\textrm{m}^{3}$).(a) Reconstruction of animals within the full imaging volume. The direction of the migration was in the positive $y$-direction. The number of swimmers and their 3D locations and orientations could be reconstructed with the scanning technique. (b)-(d) Images of a single reconstructed swimmer from the migration shown from different viewing angles. The 3D morphology of the swimmer was evident in the reconstruction, although some features of the shrimp were elongated along the scanning dimension. (e) A reference picture of \textit{Artemia salina}. }
        \label{fig:bodyReconstruction}

\end{figure*}
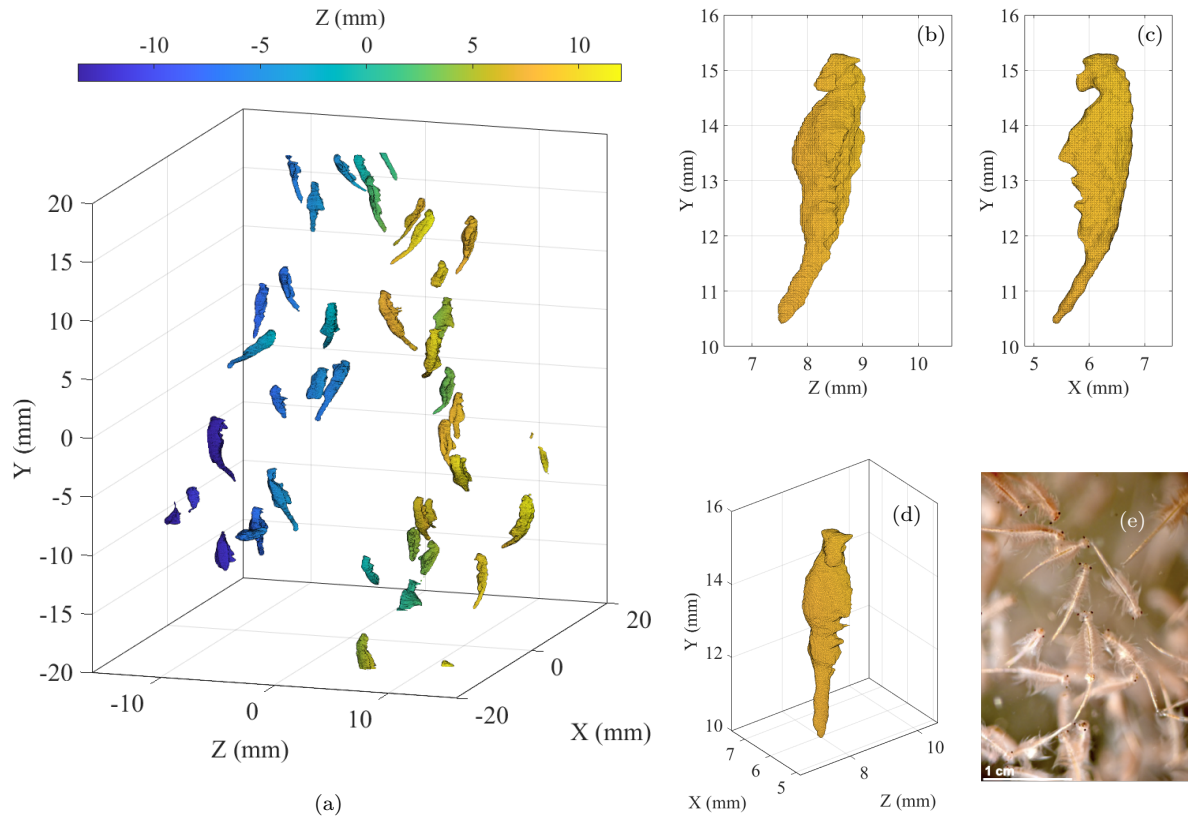

The ability of the technique to reconstruct configurations of brine shrimp during a vertical migration is illustrated in Figure \ref{fig:bodyReconstruction}, which shows a scanned reconstruction of the animals within the full imaging volume. The approximately 40 shrimp bodies are reproduced from a scan conducted approximately four minutes into the migration and represent one of the densest collections of animals imaged during the measurement campaign. Figure \ref{fig:3dmig} shows all of the reconstructed shrimp visualized within the imaging volume. The shrimp coloring indicates their depth-wise location with positive values corresponding to locations closer to the camera. These segmented images have been corrected to account for the camera perspective and deblurred along the scanning dimension to compensate for the finite sheet thickness using Richardson–Lucy deconvolution \citep{Biggs1997}. 
Despite the deblurring process, some elongation of the bodies in the scanning dimension was still evident. Figures \ref{fig:bodyReconstruction}b-\ref{fig:bodyReconstruction}d show renderings of one animal in the migration from different viewing angles. This elongation from the scanning was most apparent in the animals tails, which appeared thicker along the scanning dimension than in the imaging plane. In the future, this effect could be mitigated through further narrowing of the laser sheet with additional optical components. \mkf{Similarly, these figures also illustrate the effect of the camera perspective on the reconstruction quality. Body morphology within the line-of-sight of the camera (seen from Figure \ref{fig:bodyReconstruction}c) was reconstructed with higher fidelity than those obscured by the shrimp body. These differences are apparent in Figures \ref{fig:bodyReconstruction}b and \ref{fig:bodyReconstruction}d where details such as the organisms legs were reconstructed on the right side of the organism (large values of z) but were absent from the left side of the organism (smaller values of z).}

Even with these limitations, the reconstructed swimmers were able to capture the 3D locations, body morphology, and orientations of the scanned organisms. Though alternative single-camera techniques, such as DDPIV \citep{Troutman2018}, can similarly track particle locations in 3D, extracting a comparable level of body-specific information is neither straightforward for isolated swimmers nor possible at the high number densities present in these aggregations. Furthermore, the average animal number density measured in this scan, $8\times 10^5$ animals per $\textrm{m}^{3}$, was at the upper bound of animal number density estimates conducted in previous laboratory experiments \citep{Houghton2018,Houghton2019}. Where previous studies had to infer the animal number density during migration from depth-averaged 2D measurements, the current system was capable of measuring this quantity directly. 

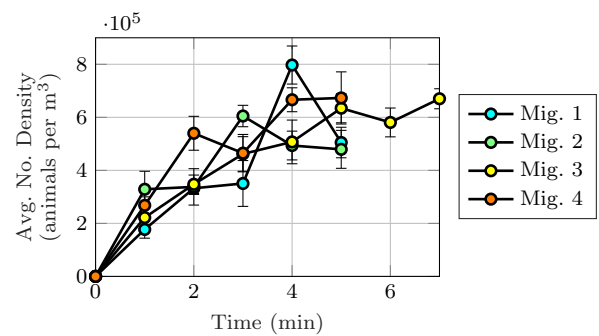
\begin{figure}[ht!]
\centering
%
%
\definecolor{mycolor1}{rgb}{0.00000,1.00000,1.00000}%
\definecolor{mycolor2}{rgb}{1.00000,1.00000,0.00000}%
\begin{tikzpicture}

\begin{axis}[%
width=1.8in,
height=1.25in, at={(0in,0in)},
scale only axis,
xmin=0,
xmax=7,
xlabel style={font=\color{white!15!black}},
xlabel={Time (min)},
ymin=0,
ymax=900000,
ylabel style={font=\color{white!15!black},align=center,text width=3cm},
ylabel={Avg. No. Density \\(animals per $\textrm{m}^3$)},
axis background/.style={fill=white},
xmajorgrids,
ymajorgrids,
legend style={at={(1.9in,0.3in)}, anchor=south west, legend cell align=left, align=left, draw=white!15!black}
]
\addplot [color=black, line width=1.0pt, mark=*, mark options={solid, fill=mycolor1}]
 plot [error bars/.cd, y dir = both, y explicit]
 table[row sep=crcr, y error plus index=2, y error minus index=3]{%
0	0	0	0\\
1	177160.436914965	32974.6330958587	32974.6330958587\\
2	332365.093186623	23111.1501830208	23111.1501830208\\
3	350535.394408671	86436.0823820365	86436.0823820365\\
4	797221.966117345	71804.3083867872	71804.3083867872\\
5	504982.954796077	57571.1560990884	57571.1560990884\\
};
\addlegendentry{Mig. 1}

\addplot [color=black, line width=1.0pt, mark=*, mark options={solid, fill=white!50!green}]
 plot [error bars/.cd, y dir = both, y explicit]
 table[row sep=crcr, y error plus index=2, y error minus index=3]{%
0	0	0	0\\
1	328579.613765363	68014.0154213687	68014.0154213687\\
2	337664.764376387	68642.1725605197	68642.1725605197\\
3	604919.611517339	40363.3655171932	40363.3655171932\\
4	493626.516532297	54244.4275558543	54244.4275558543\\
5	479241.694731509	71567.3092412722	71567.3092412722\\
};
\addlegendentry{Mig. 2}

\addplot [color=black, line width=1.0pt, mark=*, mark options={solid, fill=mycolor2}]
 plot [error bars/.cd, y dir = both, y explicit]
 table[row sep=crcr, y error plus index=2, y error minus index=3]{%
0	0	0	0\\
1	221829.094085833	35791.983179953	35791.983179953\\
2	347507.010871663	34735.4403573695	34735.4403573695\\
3	461828.489393713	65244.9753601015	65244.9753601015\\
4	507254.242448833	82396.8011053516	82396.8011053516\\
5	634446.351003167	55029.9871065978	55029.9871065978\\
6	580692.543221276	54359.6953791022	54359.6953791022\\
7	670029.857563011	37729.8185103721	37729.8185103721\\
};
\addlegendentry{Mig. 3}

\addplot [color=black, line width=1.0pt, mark=*, mark options={solid, fill=orange}]
 plot [error bars/.cd, y dir = both, y explicit]
 table[row sep=crcr, y error plus index=2, y error minus index=3]{%
0	0	0	0\\
1	268011.943025204	32132.2767456668	32132.2767456668\\
2	539809.365471668	63948.3896839925	63948.3896839925\\
3	464856.872930721	70493.0672768474	70493.0672768474\\
4	666244.378141751	45125.963306692	45125.963306692\\
5	673058.241100018	98392.1764569028	98392.1764569028\\
};
\addlegendentry{Mig. 4}

\end{axis}

\end{tikzpicture}%
    \caption{Plot of the mean animal number density in the imaging volume over the duration of four induced vertical migrations. The upper-most point corresponds to the visualization shown in Figure \ref{fig:bodyReconstruction}.}
        \label{fig:animaldensity}
\end{figure}

With the individual organisms identified, we tabulated the number of shrimp in each frame to observe the spatial and temporal evolution of the animal number density. A plot of the average number density in the imaging volume throughout four different migrations is shown in Figure \ref{fig:animaldensity}. Reconstructions from Figure \ref{fig:bodyReconstruction} correspond to the fourth minute of the first migration. Unlike previous experiments, specifically \cite{Houghton2019}, where a steady-state saturation in the number of shrimp was observed after one to two minutes, we observed a slow but continual growth in the number of shrimp in the frame, even up to four minutes. This slower migratory behavior may be attributed to differences in the age and health condition of the organisms tested presently or due to natural biological variability in the migratory behavior. For the present purposes, it is sufficient to note that the repeated measurements are qualitatively consistent.

\begin{figure}[ht!]
\centering
 {\includegraphics[trim={0cm 0pt 0cm 0pt},clip,width = 0.45\textwidth]{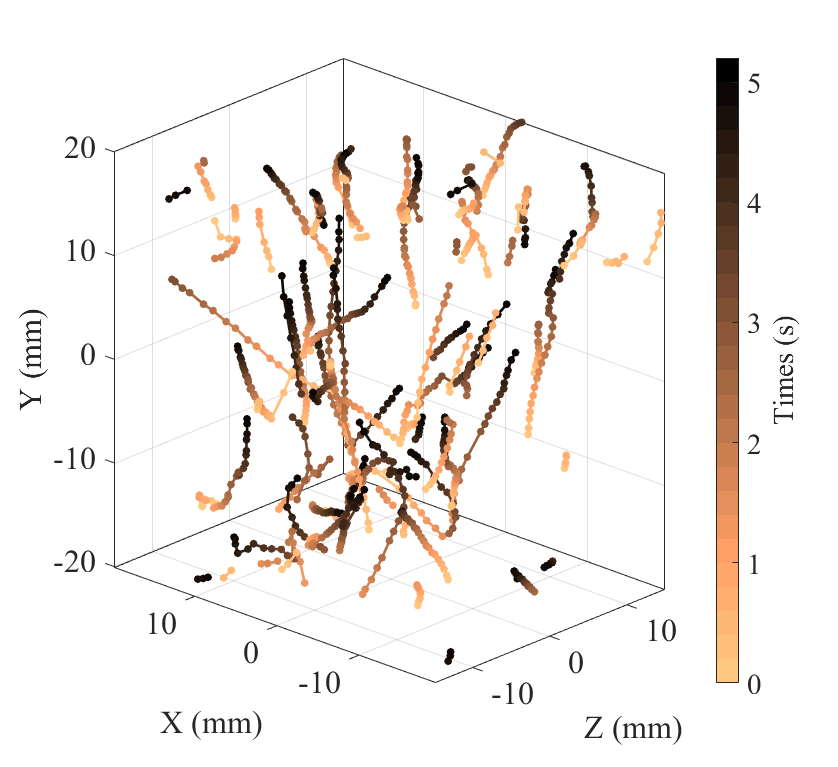}}
    \caption{Plot animal trajectories from Figure \ref{fig:bodyReconstruction} over the 5 sec. duration of the scan indicated by the color change from copper to black.}
        \label{fig:traces}
\end{figure}


Figure \ref{fig:traces} shows the animal pathlines over the scanning period associated with the swimmers reconstructed in Figure \ref{fig:bodyReconstruction}. For the scanning frequency used here ($f_s=5 $ Hz), the displacement of individual shrimp between frames is typically a fraction of a body length. Because the displacement of each organism between frames is small relative to the inter-organism spacing, we can successfully track most organisms in these experiments with a nearest-neighbor search. More sophisticated particle tracking algorithms such as that of \cite{Ouellette2006} could improve the trajectory length and prediction.

\subsection{Velocity Measurements}
\label{sec:results:velocity}


\begin{figure}[ht!]
\centering


\begin{subfigure}[b]{0.48\textwidth}
\includegraphics[trim={0 0 0 1cm},clip,width = \textwidth]{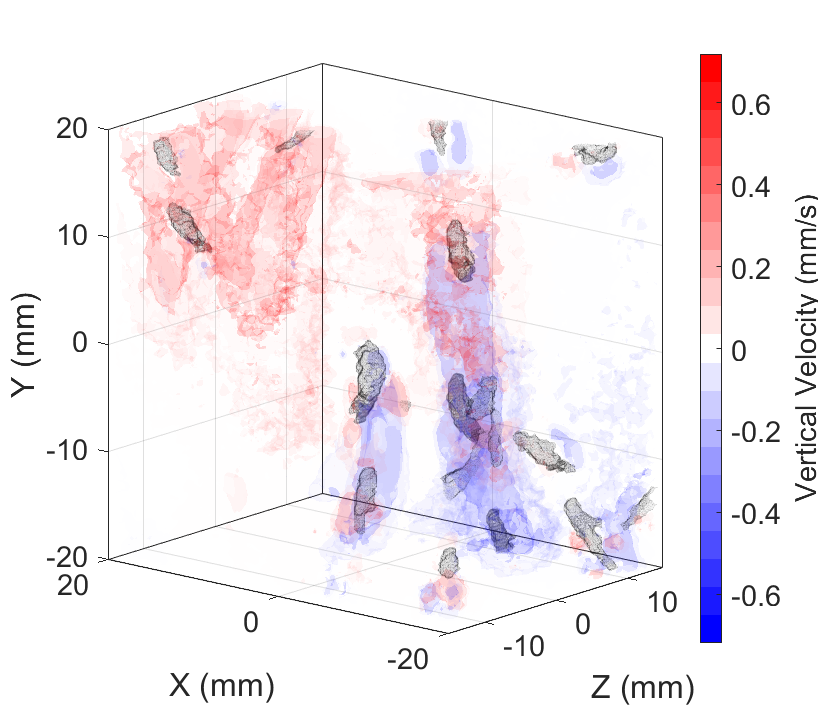}
    \caption{}
        \label{fig:23_2}
\end{subfigure}

\begin{subfigure}[b]{0.48\textwidth}
\includegraphics[width = \textwidth]{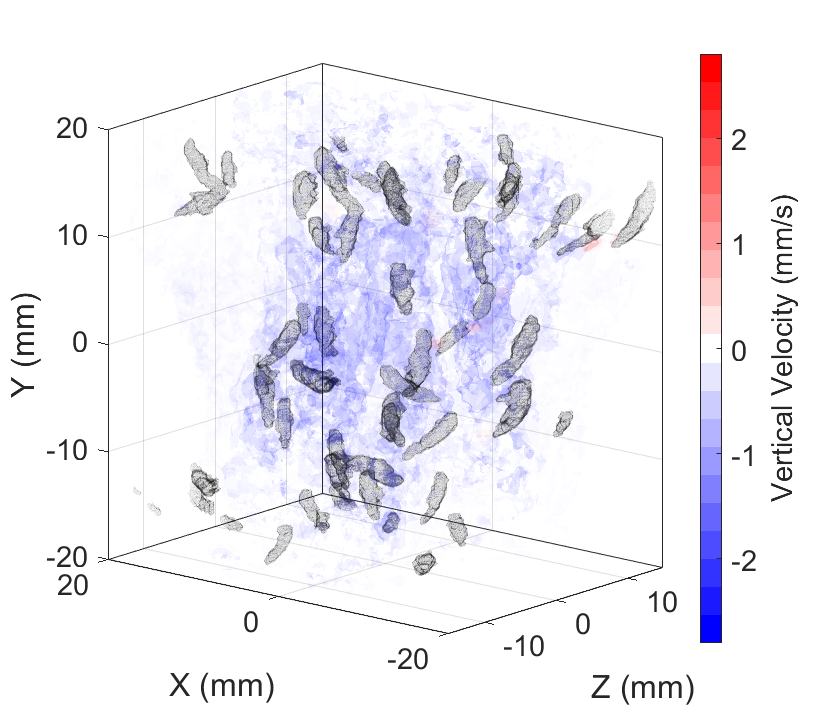}
        \caption{}
        \label{fig:23_4}
\end{subfigure}
\caption{Vertical velocity contours from snapshots taken during a single migration. Swimmer masks within the PIV field are shown in tan. Blue and red isosurfaces denote contours of negative and positive vertical velocity, respectfully. (a) Vertical velocity field 2 minutes into the migration. Downward velocity (blue) associated with the wakes of the swimming was resolved. (b) Vertical velocity field 4 minutes into the migration. Downward velocity (blue) contours associated with Figure \ref{fig:3dmig}. Instead of individual wakes, a coherent downward motion was present through the migration. }
        \label{fig:velocity}
\end{figure}

Corresponding contours of vertical velocity associated with the upward migration are shown in Figure \ref{fig:velocity}. Figure \ref{fig:23_2} shows contours of the vertical velocity contours taken from a scan taken approximately 2 minutes into the migration. This scan was obtained closer to the beginning of the migration and contained fewer animals within the imaging volume than Figure \ref{fig:3dmig}. Consequently, the technique was able to resolve downward projecting wakes from the individual swimmers. Figure \ref{fig:23_4} shows contours of the vertical velocity associated with the scan shown in Figure \ref{fig:3dmig} where the downward velocity was largest. A coherent downward motion of fluid through the aggregation was evident. This behavior was consistent with the observations of \citet{Houghton2018}, who qualitatively visualized a similar coherent downward flow from vertical migration \textit{A. salina} using planar laser-induced fluorescence. These measurements indicate that the technique was capable of quantitatively resolving the 3D velocity field in and around the swimming aggregation.


\section{Velocity Measurement Validation}
\label{sec:jet}

\begin{figure*}[ht]
    \centering
            \includegraphics[trim = {0 3.5cm 0cm 0 },clip,width = \textwidth]{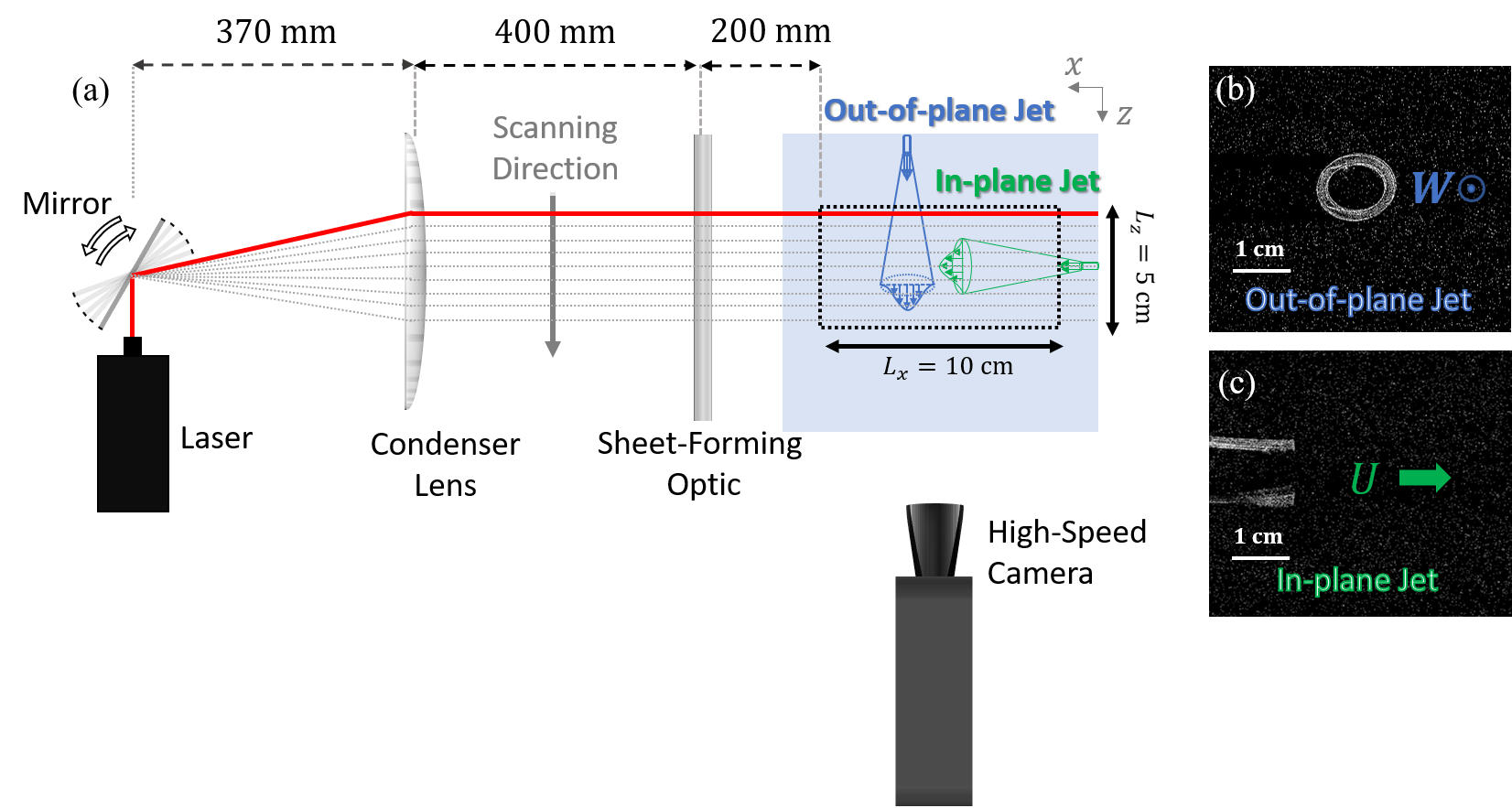}
       \caption{Diagrams of the scanning system setup for flow velocity measurement validation. (a) Top view of the experimental setup. From this viewpoint, the scanning direction was parallel with the page's height, and the imaging planes were aligned with the page width and normal directions. The two orientations of the jet flow measured in this experiment corresponding to flow normal and parallel to the imaging plane are blue and green, respectively. (b) Image of tracer field and jet outlet in the out-of-plane orientation. The optical axis of the camera was parallel to the axis of the jet, and flow was out of the page. The elliptical cross-section seen in the image was due to plastic deformation of the tube prior to installation and was present in both configurations. (c) Image of tracer field and jet outlet as imaged from the in-plane orientation. The optical axis of the camera was perpendicular to the axis of the jet, and the flow advection was from left to right.}
               \label{fig:jetflow}
\end{figure*}
Due to the lack of a ground truth reference to validate the velocity measurements in the vertical migrations, this capability of the measurement system was assessed by using a controlled laminar jet flow.
By evaluating the system against a laminar jet flow without an aggregation present, we were able to ensure that it could accurately resolve the three-component, three-dimensional velocity field. 
Imaging was conducted in a small $40\, \times40\,\times40\,\textrm{cm}^3$ glass tank seeded with $100\,\mu \textrm{m}$ silver-coated glass spheres. A syringe pump provided a bulk flow of $21.50\, \textrm{mL}/ \textrm{min}$ into a length of Tygon tubing with an elliptical cross section (equivalent diameter, $D_e=8.7$ mm) which exited as a laminar free jet flow of $Re_D = U_b D_e /\nu$, where $U_b$ was the bulk jet velocity, and $\nu$ was the water kinematic viscosity ($0.95$ cSt at $22^\circ$ C). The finite eccentricity of the cross-section was due to plastic deformation of the tube wall prior to installation.  Illumination was provided by a 532 nm laser with Gaussian beam shape. Here, the scanning speed, $u_s$, was $100$ times larger than the jet's bulk velocity, which was sufficient to resolve the 3D particle positions with minimal error related to the finite scanning speed \citep{Kozul2019}. 
Imaging parameters and jet specifications are listen in Table \ref{tab:imaging_params_jet}. 

Measurements of the out-of-plane velocity component aligned with the scanning direction were verified by scanning the jet in two different orientations, as shown in Figure \ref{fig:jetflow}. In the first orientation (given by the blue jet in Figure \ref{fig:jetflow}a), the axis of the jet was parallel to the scanning direction such that the jet flow was normal to the imaging plane. An image slice of the jet in this configuration can be seen in Figure \ref{fig:jetflow}b with the elliptical cross-section of the wall illuminated by the imaging sheet. Correspondingly, the tracer particle motion was primarily out of the page. In the second orientation (given by the green jet in Figure \ref{fig:jetflow}a), the jet's axis was perpendicular to the scanning direction, and the tracer particle displacements were primarily contained within the same imaging plane. In the corresponding image slice (see Figure \ref{fig:jetflow}c), the imaged cross-section of the tube instead appeared rectangular, and the fluid advection was from left to right. Resolving the flow in the first configuration depended on the ability of the technique to reconstruct tracer particle location along the scanning dimension. There, the velocity calculations correlated particles across different image sheets. Conversely, in the second configuration, velocity calculations were far less sensitive to the scanning effect as fluid motion was primarily contained within the image plane. Consequently, the fluid motion could still be determined without explicitly relying on the particles motions in adjacent image sheets similar to conventional 2D PIV. Hence, the in-plane jet measurement provided a ground truth reference for the out-of-plane measurements. 

\begin{table}[ht]
\centering
\begin{scriptsize}
\begin{tabular}{lr|c|c}
\hline
& Parameter    &     Symbol         &        Value           \\
\hline \hline
\parbox[c]{1mm}{\multirow{8}{*}{\rotatebox[origin=t]{90}{Scanning Parameters}}}
&Field of View       &   $L_x \!\times\! L_y$  &   $100\,\times100\,\textrm{mm}^2\,$ \\
&Depth of Field      &   $L_z$    &   $47\,\textrm{mm}$ \\
&Approx. Voxel Size  &   \rule{0.5cm}{0.5pt}     &   $100\times100\times100\,\mu\textrm{m}^3$ \\
&Image Acq. Rate    &   $f_c$     &   $5,\!000$ fps\\
&Scanning Freq.    &   $f_s$     &   $10$ Hz\\
&Scanning Speed    &   $u_s$     &   $50\;\textrm{cm} \cdot \textrm{s}^{-1}$ \\
&$e^{-2}$ Sheet thickness    &   $h$     &   $\approx1\;\textrm{mm}$ \\
&Sheet step size    &   $\Updelta z$    &   $100\,\mu\textrm{m}$ \\
\hline
\parbox[c]{1mm}{\multirow{3}{*}{\rotatebox[origin=c]{90}{Jet }}} & Bulk Velocity    &   $U_b$    &   $5.4 \pm 0.1\,\textrm{mm/s}$ \\
& Equivalent Diameter &   $D_e$    &   $8.7\pm0.2\,\textrm{mm}$ \\
& Aspect Ratio &   \rule{0.5cm}{0.5pt}    &   $1.31$ \\
\hline
& Reynolds Number &   $Re_D$    &   $50$ \\
& Non-dim Sheet Speed    &   $u_s/U_b$    &   92 \\
& Non-dim Sheet Width    &   $h/\Updelta z$    &   $10$ \\
& No. laser sheets  &   $N$     &   $475$ \\
& No. of scans  &   \rule{0.5cm}{0.5pt}    &   $40$ \\
\hline
\end{tabular}
\end{scriptsize}
\caption{Imaging parameters for jet measurements}
\label{tab:imaging_params_jet}
\end{table}

\begin{figure}
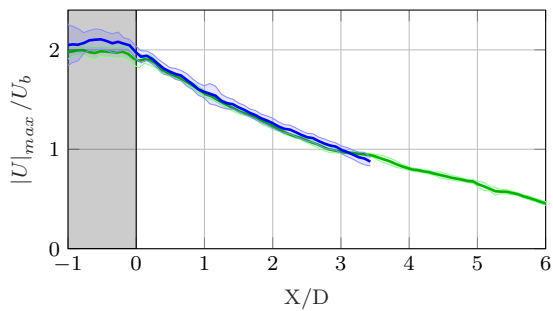

\centering
    \include{Figures/JetPics/jetvelocity}
        \caption{Maximum fluid velocity ($U_{\textrm{max}}$) exiting the round jet as a function of distance from the jet exit. The gray area denotes measurements obtained inside the jet tube. (\myline{blue}): Out-of-plane jet (blue) where the fluid velocity was normal to the imaging plane. (\myline{black!30!green}): In-plane jet (green) where the velocity was parallel to the imaging plane. Blue and green bands correspond to the standard deviation of the local velocity measurements in the out-of-plane and in-plane configurations, respectively.}
        \label{fig:jetplot}
\end{figure}

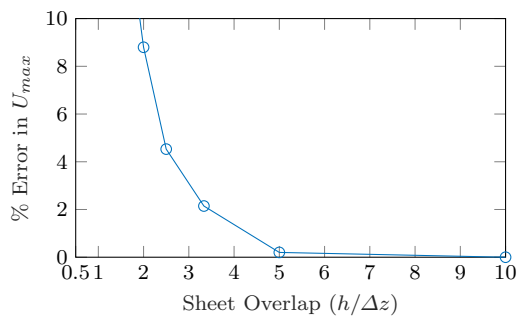
\begin{figure}
\centering
%
%
\definecolor{mycolor1}{rgb}{0.00000,0.44700,0.74100}%
\begin{tikzpicture}

\begin{axis}[%
width=2.25in,
height=1.25in,
at={(0in,0in)},
scale only axis,
xmin=0.5,
xmax=10,
xminorticks=true,
xlabel style={font=\color{white!15!black}},
xlabel={Sheet Overlap $(h/\Delta z)$},
ymin=0,
ymax=10,
yminorticks=true,
xtick={0.5,1,2,3,4,5,6,7,8,9,10},
xticklabels={0.5,1,2,3,4,5,6,7,8,9,10},
ylabel style={font=\color{white!15!black}},
ylabel={$\%$ Error in $U_{max}$},
axis background/.style={fill=white}
]
\addplot [color=mycolor1, mark=o, mark options={solid, mycolor1}, forget plot]
  table[row sep=crcr]{%
10 0 \\
5	0.201095104279633\\
3.33333333333333	2.14515224489383\\
2.5	4.53005171347735\\
2	8.79834053698205\\
1.66666666666667	13.576167970865\\
1.25	21.6376340218176\\
1	40.4823583090189\\
0.833333333333333	66.2477084454481\\
0.625	89.9801762836379\\
0.5	95.1040442191583\\
};
\end{axis}


\end{tikzpicture}%
        \caption{Error in the max jet velocity calculation relative to $h/\Updelta z = 10$ as a function of sheet overlap. }
        \label{fig:skipplot}
\end{figure}
The two different configurations were evaluated by reorienting the jet within the tank while keeping the imaging system fixed. To evaluate the technique, we compared the maximum fluid velocities from each configuration as a function of distance from the jet exit, as shown in Figure \ref{fig:jetplot}. Because the jet tube is translucent, the technique is capable of measuring velocity inside the jet tube (shaded in gray), albeit with a slight difference between the two orientations. In this region, the out-of-plane orientation measures approximately $5\%$ larger than its in-plane counterpart. Immediately outside the jet exit, there is excellent agreement between the two measurements over the extent of the domain. The distance over which data is reported for the out-of-plane jet is considerably shorter than the in-plane jet due to the depth of field being smaller than the image width. Importantly, this test indicates that the setup is capable of resolving velocities both parallel and normal to the imaging plane and is consistent with previously reported validations of scanning PIV \citep{Brucker2013,Kozul2019}. 

Here, because of the fine sampling of the imaging volume, we were able to examine how the quality of the velocity calculations along the optical axis degraded with increased sheet spacing. For the experiment conducted here, the step size between consecutive images was approximately the voxel size which corresponded to a $90\%$ overlap between adjacent sheets (i.e., $h / \Delta z = 10$). Here, we artificially increased the step size of our data set by first down-sampling the image volumes from full-resolution scans and then re-interpolating via tri-spline the new images back to the full resolution. The images were then processed with the same cross - correlation algorithm and compared with the full-resolution result ($h / \Delta z = 10$). Figure \ref{fig:skipplot} shows how the mean difference between the max velocity calculation shown in Figure \ref{fig:jetplot} varied as the effective step size between laser sheets was increased. Importantly, even when the data was down-sampled by a factor of 2 ($h / \Delta z = 5$), there was a negligible change in the measured maximum jet velocity over the domain. \mkf{Even downsampling the data by a factor of 3 ($h / \Delta z = 3.33$), yielded a mean error of approximately $2\%$ over the imaging domain compared to the full resolution measurement. Above this range, the error began to increase sharply as the spacing approached the sheet width. These empirical results were consistent with previous findings from comparable numerical investigations including \citet{Kozul2019} who found a sheet overlap of $h / \Delta z = 5$ to be sufficiently resolved for particle tracking and \citet{Lawson2014} who found $h / \Delta z = 3-4$ to be optimal for single camera measurements.} 


\section{Discussion and Conclusions}
\label{sec:discussion}

\label{sec:conclusion}

A 3D scanning velocimetry system for 3D-3C velocity measurements and particle aggregation reconstruction was demonstrated using an induced vertical migration of \textit{A. salina}. The technique successfully reconstructed the swimmer bodies and their 3D configurations at animal number densities at the upper bound of those found in previous laboratory migration experiments ($8 \times 10^5$ animal per $\textrm{m}^{3}$), a task that had not been accomplished with previous methods. This capability will allow for more direct studies of the flow-structure interactions that enable individual animal wakes to coalesce into larger-scale flows. \mkf{The success of this technique at these animal number densities suggest that it could have broader applications in the study of flows with dispersed particles. The animal volume fractions measured in this study, $\Phi_v \leq 1.7 \pm 0.24 \times 10^{-3}$, encompass the range of volume fractions ($10^{-6} \leq \Phi_v \leq 10^{-3}$) over which two-way coupling is exhibited between turbulence and dispersed particles \citep{Elghobashi1994}. This capability suggests that scanning techniques could be a robust tool for studying this coupling in turbulent flows with translucent or transparent particles, such as bubbles and droplets.}

The most notable challenges for this system included the trade-offs between the temporal resolution of the flow field, illumination of the images, and the resolvable depth of field. The achievable depth of field in the present design was primarily constrained by the power of the laser. Increasing the depth of field to keep all of the scans in sharp focus required significantly reducing the image illumination due to compounding effects of shrinking the lens aperture and increasing the camera frame rate. In the case of the former, reducing the aperture caused a quadratic reduction in the light intensity for a linear increase in the depth of field. 

Future implementations of this technique can employ a telecentric lens on the high-speed camera to ensure a constant magnification throughout the entire image volume, eliminating any parallax. Additionally, this lens type will also allow for a larger usable depth of field for a given aperture due to the symmetric image blurring. Similarly, incorporating a scanning lens into the setup could significantly improve the temporal capabilities of the scanning system. A scanning lens would allow the location of the focal plane to be adjusted over distances comparable to the field of view at bandwidths exceeding the scanning frequencies. By synchronizing the focal distance to the laser sheet location, the depth of field can be reduced to the thickness of the laser sheet and individual images can be captured by the high-speed camera using a much larger aperture. This modification would allow for significantly greater illumination of the camera sensor than the current implementation where the focal plane is static, and the entire scanning distance must be contained within depth of field. 

Lastly, because the technique relies on a single high-speed camera, it is compatible with many existing underwater imaging systems such as the diver operated self‐contained underwater velocimetry apparatus (SCUVA) \citep{Katija2008} or remotely operated DeepPIV \citep{Katija2020a,Katija2017a}. Adapting this technique for field deployment could enable 3D-3C velocity measurements of various environmental and biological flows that have traditionally been limited to 2D observations. Similar to the \textit{A. salina}, there are numerous marine organism whose feeding and swimming are potentially observable with this technique, including salps, jellyfish, siphonophores, and ctenophores. The ability to image the 3D flow in and around these organisms could provide numerous biological and fluid mechanical insights.


\begin{acknowledgements}
    The authors would like to thank Prof. Christian Franck for supplying the basis for the cross-correlation algorithm used in this study. This work was supported by the U.S. National Science Foundation Grant, under Award Number 1510607 and the Gordon and Betty Moore Foundation.
 \end{acknowledgements}

\bibliographystyle{spbasic}      
\bibliography{references.bib}
\newpage



 
\end{document}